# The Next Generation Very Large Array: A Technical Overview

Robert J. Selina*, Eric J. Murphy, Mark McKinnon, Anthony Beasley, Bryan Butler, Chris Carilli, Barry Clark, Alan Erickson, Wes Grammer, James Jackson, Brian Kent, Brian Mason, Matthew Morgan, Omar Ojeda, William Shillue, Silver Sturgis, Denis Urbain.

National Radio Astronomy Observatory, Socorro, NM USA 87801

## ABSTRACT

The next-generation Very Large Array (ngVLA) is an astronomical observatory planned to operate at centimeter wavelengths (25 to 0.26 centimeters, corresponding to a frequency range extending from 1.2 GHz to 116 GHz). The observatory will be a synthesis radio telescope constituted of approximately 214 reflector antennas each of 18 meters diameter, operating in a phased or interferometric mode.

We provide an overview of the current system design of the ngVLA. The concepts for major system elements such as the antenna, receiving electronics, and central signal processing are presented. We also describe the major development activities that are presently underway to advance the design.

## 1. INTRODUCTION

As part of its mandate as a national observatory, the National Science Foundation's (NSF) National Radio Astronomy Observatory (NRAO) is looking toward the long-range future of radio astronomy and fostering the long-term growth of the US and global astronomical community. With NSF support, NRAO has sponsored a series of science and technical community meetings to define the science mission and concept for a next-generation Very Large Array (ngVLA) [1], building on the legacies of the Atacama Large Millimeter/submillimeter Array (ALMA) and the Jansky Very Large Array (JVLA).

Based on input solicited from the astronomical community, the ngVLA is planned as an astronomical observatory that will operate at centimeter wavelengths (25 to 0.26 centimeters, corresponding to a frequency range extending from 1.2 GHz to 116 GHz). The observatory will be a synthesis radio telescope with a main array constituted of approximately 214 reflector antennas each of 18 meters diameter, operating in a phased or interferometric mode.

A dense core short baseline array (SBA) of 19 reflector antennas of 6m aperture will be sensitive to a portion of the larger angular scales undetected by the main array. The SBA may be combined with 4 18m (main-array) antennas used in total power mode to completely fill in the central hole in the *(u, v)*-plane left by the 6m dishes.

The ngVLA will have approximately ten times the sensitivity of the JVLA and ALMA, with more than thirty times longer baselines (~1000km) providing milliarcsecond-resolution, plus a dense core on km-scales for high surface brightness sensitivity. Such an array bridges the gap between ALMA, a superb sub-mm array, and the future SKA1, optimized for longer wavelengths.

The dense core and the signal processing center of the array will be located at the Very Large Array site, on the plains of San Agustin, New Mexico. The array will include stations in other locations throughout the state of New Mexico, Arizona, west Texas, and northern Mexico. The high desert plains of the Southwest US, at over 2000m elevation, provide excellent observing conditions for the frequencies under consideration, including reasonable phase stability and opacity at 3mm wavelength over a substantial fraction of the year.

Operations will be conducted from both the VLA site facilities and the Array Operations Center in Socorro, NM. Additional operations centers may be incorporated into the design.

The facility will be operated as a PI-driven instrument. The fundamental data products for ngVLA users will be science-ready data products (i.e., images and cubes) generated using calibration and imaging pipelines created and maintained by the project. Both the pipeline products and the "raw" visibilities and calibration tables will be archived, retaining the option of future re-processing and archival science projects.

*rselina@nrao.edu;    phone 1 575 835-7000;    fax 1 575 835-7026;    ngvla.nrao.edu

The ngVLA project is developing a Reference Design for the array that will form a baseline for construction and operation costing, and future design trade-off decisions. This Reference Design is intended to be low technical risk in order to provide a degree of conservativism in the estimates. However, leading-edge concepts and techniques that may help improve the performance and/or reduce cost are being developed in parallel, and will be evaluated in the conceptual design phase of the facility.

## 2. TELESCOPE SPECIFICATIONS

### 2.1 Reference Array Performance

The predicted performance of the array is summarized in Table 1. This is an update to the performance estimates originally documented in [2].

Table 1 - ngVLA Key Performance Metrics

| Center Frequency | 2.4 GHz | 8 GHz | 16 GHz | 27 GHz | 41 GHz | 93 GHz | Notes |
|---|---|---|---|---|---|---|---|
| Band Lower Frequency [GHz] | 1.2 | 3.5 | 12.3 | 20.5 | 30.5 | 70.0 | a |
| Band Upper Frequency [GHz] | 3.5 | 12.3 | 20.5 | 34.0 | 50.5 | 116.0 | a |
| Field of View FWHM [arcmin] | 24.4 | 7.3 | 3.7 | 2.2 | 1.4 | 0.6 | b |
| Aperture Efficiency | 0.78 | 0.77 | 0.86 | 0.85 | 0.81 | 0.60 | b |
| Effective Area, $A_{eff}$ x $10^3$ [m$^2$] | 42.2 | 41.7 | 46.8 | 46.0 | 44.0 | 32.4 | b |
| System Temp, $T_{sys}$ [K] | 23 | 25 | 22 | 33 | 45 | 62 | a, e |
| Max Inst. Bandwidth [GHz] | 2.3 | 8.8 | 8.2 | 13.5 | 20.0 | 20.0 | a |
| Sampler Resolution [Bits] | 8 | 8 | 8 | 4 | 4 | 4 | |
| Antenna SEFD [Jy] | 328.6 | 361.8 | 283.2 | 432.4 | 617.0 | 1153.7 | a, b |
| Resolution of Max. Baseline [mas] | 26 | 8 | 4 | 2.3 | 1.5 | 0.7 | c |
| Resolution FWHM @ Natural Weighting [mas] | 163 | 49 | 24 | 14 | 10 | 4 | c, d |
| Continuum rms, 1 hr [$\mu$Jy/beam] | 0.41 | 0.23 | 0.19 | 0.22 | 0.26 | 0.48 | d |
| Line Width, 10 km/s [kHz] | 80.0 | 266.7 | 533.3 | 900.0 | 1366.7 | 3100.0 | |
| Line rms, 1 hr, 10 km/s [$\mu$Jy/beam] | 69.0 | 41.6 | 23.0 | 27.1 | 31.3 | 38.9 | d |

(a) 6-band 'baseline' receiver configuration.
(b) Reference design concept of 214 18m aperture antennas. Unblocked aperture with 160um surface.
(c) Rev. B 2018 Configuration. Resolution in EW axis.
(d) Point source sensitivity using natural weights, dual pol, and all baselines.
(e) At the nominal mid-band frequency shown. Assumes 1mm PWV for W-band, 6mm PWV for others; 45 deg elev. on sky for all.

The continuum and line rms values in Table 1 are for point source sensitivity with a naturally weighted beam. Imaging sensitivity is estimated based on [3] and provided as a function of angular resolution in Table 2. The table is by necessity a simplification and the imaging sensitivity will vary from these reported values depending on the quality (defined as the ratio of the power in the main beam attenuation pattern to the power in the entire beam attenuation pattern as a function of the FWHM of the synthesized beam [4]) of the (sculpted) synthesized beam required to support the science use case.

The brightness sensitivity of an array is critically dependent on the array configuration. The ngVLA has the competing desires of both good point source sensitivity at full resolution, and good surface brightness sensitivity on scales similar to the primary beam size. Different array configurations that might provide a reasonable compromise through judicious weighting of the visibilities for a given application have been explored [5] (See [6] for similar studies for the SKA). It is important to recognize the fact that for any given observation, from full resolution imaging of small fields, to imaging structure on scales approaching that of the primary beam, some compromise will have to be accepted.

Table 2 – Projected imaging sensitivity as a function of angular resolution.

| Center Frequency, f | 2.4 GHz | 8 GHz | 16 GHz | 27 GHz | 41 GHz | 93 GHz |
|---|---|---|---|---|---|---|
| **Resolution [mas]** | 1000 | | | | | |
| Continuum rms, 1 hr, Robust [$\mu$Jy/beam] | 0.80 | 0.50 | 0.43 | 0.53 | 0.64 | 1.28 |
| Line rms 1 hr, 10 km/s Robust [$\mu$Jy/beam] | 135.6 | 91.0 | 53.3 | 65.2 | 77.9 | 102.6 |
| Brightness Temp ($T_B$) rms continuum, 1 hr, Robust [K] | 0.1688 | 0.0095 | 0.002 | 0.0009 | 0.0005 | 0.0002 |
| $T_B$ rms line, 1 hr, 10 km/s, Robust [K] | 28.62 | 1.73 | 0.25 | 0.11 | 0.06 | 0.01 |
| **Resolution [mas]** | 100 | | | | | |
| Continuum rms, 1 hr, Robust [$\mu$Jy/beam] | 0.63 | 0.40 | 0.35 | 0.44 | 0.53 | 1.07 |
| Line rms 1 hr, 10 km/s Robust [$\mu$Jy/beam] | 106.4 | 73.4 | 43.5 | 53.8 | 64.7 | 86.1 |
| Brightness Temp ($T_B$) rms continuum, 1 hr, Robust [K] | 13.246 | 0.767 | 0.167 | 0.073 | 0.039 | 0.015 |
| $T_B$ rms line, 1 hr, 10 km/s, Robust [K] | 2245.9 | 139.4 | 20.7 | 9.0 | 4.7 | 1.2 |
| **Resolution [mas]** | 10 | | | | | |
| Continuum rms, 1 hr, Robust [$\mu$Jy/beam] | - | 0.31 | 0.27 | 0.35 | 0.43 | 0.87 |
| Line rms 1 hr, 10 km/s Robust [$\mu$Jy/beam] | - | 55.8 | 33.8 | 42.3 | 51.4 | 69.7 |
| Brightness Temp ($T_B$) rms continuum, 1 hr, Robust [K] | - | 58.3 | 12.9 | 5.8 | 3.1 | 1.2 |
| $T_B$ rms line, 1 hr, 10 km/s, Robust [K] | - | 10596 | 1605 | 706 | 372 | 98 |

### 2.2 Community-Led Science Options

Three community-led options have been identified that could expand the capabilities of the ngVLA beyond the reference design and may be subsumed into the overall ngVLA concept:

- A VLBI expansion. [7]
- A commensal low-frequency (<1 GHz) system and aperture array. [8]
- A pulsar timing telescope, re-using the VLA 25m antennas. [18]

These options are not included in the baseline design or the performance estimates given above, and funding for the construction and operation of these options would need to be separately secured.

The science white papers present a number of compelling VLBI astrometric science programs made possible by the increased sensitivity of the ngVLA. These astrometric programs require excellent sensitivity per baseline, but may not require dense coverage of the *(u, v)*-plane, since high dynamic range imaging may not be required.

Within the scope of the baseline design, the ngVLA can be used as an ultra-sensitive, anchoring instrument, in concert with radio telescopes across the globe. Such a model would parallel the planned implementation for sub-mm VLBI, which employs the ultra-sensitive phased ALMA, plus single dish sub-mm telescopes around the globe, to perform high priority science programs, such as imaging the event horizons of supermassive black holes [9]. A VLBI expansion option would include out-lying stations (likely at the VLBA sites and Green Bank, WV) within the ngVLA construction plan itself, perhaps comprising up to 10% of the total area, out to trans-continental baselines.

The commensal low-frequency option relies on installing a prime focus warm receiver system to observe the 150 MHz to 800 MHz bands concurrently with high frequency observations. An ancillary aperture array accesses frequencies below 150 MHz while leveraging common infrastructure deployed for the ngVLA. The scientific opportunities enabled by such a system include a continuous synoptic survey for transients, complementary low frequency images of all high-frequency targets and their environments, and space weather applications.

The pulsar timing telescope option would provide GBT-scale sensitivity as a phased array, with 80%+ dedicated time to pulsar timing. Such an option affordably supports existing pulsar timing projects such as NanoGRAV [10] ensuring they continue to receive large time allocations for timing of PTAs. The system could employ prime focus feeds, ngVLA electronics, and a custom back-end instrument to reduce the operations burden associated with operating the 28-element phased array.

## 3. NEW PARAMETER SPACE

Figure 1 shows a slice through the parameter space, resolution versus frequency, covered by the ngVLA along with other existing and planned facilities that are expected in the 2030s at all wavelengths. The maximum baselines of the ngVLA imply a resolution of better than 3mas at 1cm. As we shall see below, coupled with the high sensitivity of the array, this resolution provides a unique window into the formation of planets in disks on scales of our own Solar system at the distance of the nearest active star forming regions.

Figure 2 shows a second slice through parameter space: effective collecting area versus frequency. A linear-linear plot highlights the parameter space opened by the ngVLA. Note that the SKA-1 will extend to below 100MHz while ALMA extends up to almost a THz.

We note that there are other aspects of telescope phase space that are relevant, including field of view, mapping speed, surface brightness sensitivity, bandwidth, system temperature, dynamic range, etc. We have presented the two principle and simplest design goals, namely, maximum spatial resolution and total effective collecting area (as a gross measure of system sensitivity).

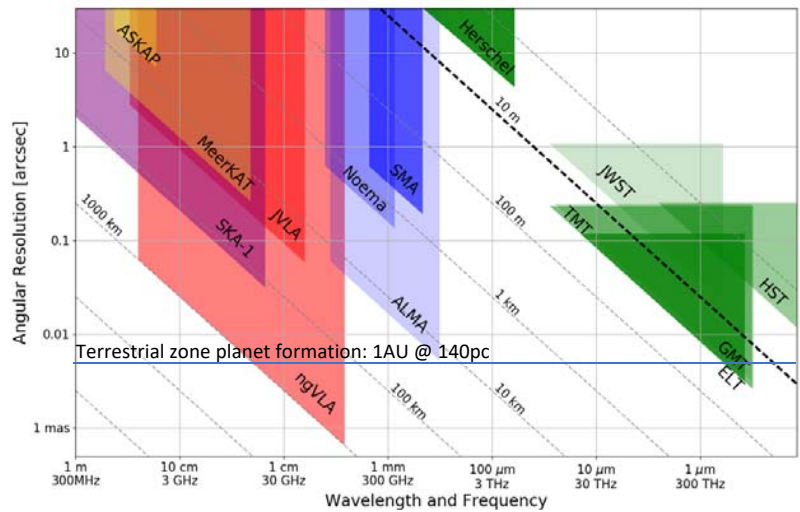

Figure 1. Spatial resolution versus frequency set by the maximum baselines of the ngVLA as compared to that of other existing and planned facilities.

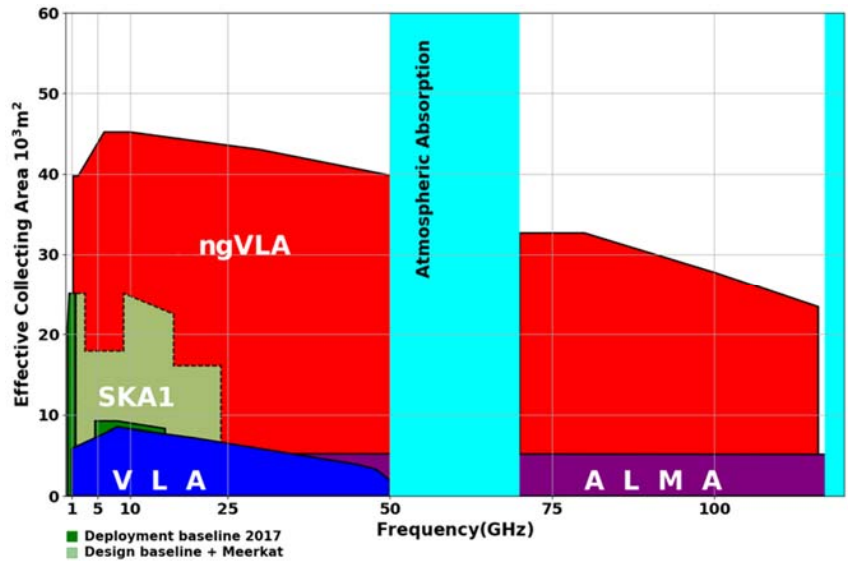

Figure 2. Effective collecting area versus frequency for the ngVLA as compared to that for other existing or planned facilities. Note that lower and higher frequencies are not shown (e.g. SKA-1 will extend to below 100 MHz and ALMA extends up to about a THz). Both the SKA1 'deployment baseline' (dark green) and 'design baseline' (light green) are shown, inclusive of the MeerKAT array.[23]

## 4. KEY SCIENCE GOALS

The project solicited input on the science goals and requirements for the facility and these were captured in the form of science use cases by the Science Working Groups. Over 80 use cases were documented by more than 200 authors, and then ranked by the Science Advisory Council. The key science goals for the facility [11] are summarized below.

### 4.1 KSG1: Unveiling the Formation of Solar System Analogues

The ngVLA will measure the planet initial mass function to 5 – 10 Earth masses and unveil the formation of planetary systems similar to our Solar System by probing the presence of planets on orbital radii as small as 0.5 AU at a distance of 140 pc. The ngVLA will also reveal circum-planetary disks and sub-structures in the distribution of millimeter-size dust particles created by close-in planets and will measure the orbital motion of these features on monthly timescales.

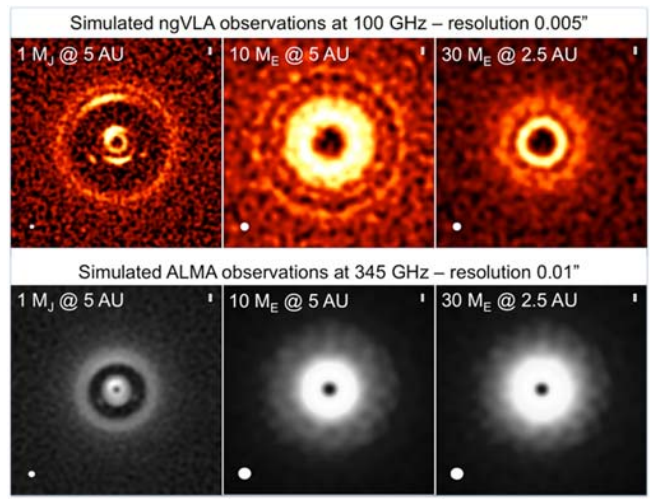

Figure 3. Ricci et al. (2018) – ngVLA- (top row) and ALMA- (bottom row) simulated observations of protoplanetary disk continuum emission perturbed by a Jupiter mass planet orbiting at 5 AU (left column), a 10 Earth mass planet orbiting at 5 AU (center column), and a 30 Earth mass planet orbiting at 2.5 AU (right column). The ngVLA

observations at 100 GHz were simulated with 5 mas angular resolution and an rms noise level of 0.5 μJy/bm. ALMA observations at 345 GHz where simulated assuming the most extended array configuration comprising baselines up to 16 km and a rms noise level of 8 μJy/bm.

### 4.2  KSG2: Probing the Initial Conditions for Planetary Systems and Life with Astrochemisty

The ngVLA will detect predicted, but as yet unobserved, complex prebiotic species that are the basis of our understanding of chemical evolution toward amino acids and other biogenic molecules. The ngVLA will enable the detection and study of chiral molecules, testing ideas on the origins of homochirality in biological systems. The detection of such complex organic molecules will provide the chemical initial conditions of forming solar systems and planets.

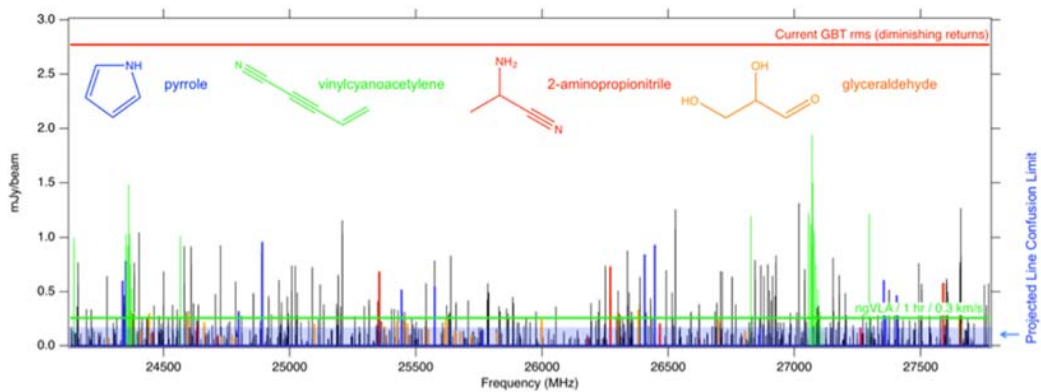

Figure 4. A conservative simulation of 30 as-yet-undetected complex interstellar molecules (black) likely to be observed by the ngVLA above the confusion limit around hot cores with typical sizes of ~1" – 4". Key molecules are highlighted in color.

### 4.3  KSG3: Charting the Assembly, Structure, and Evolution of Galaxies from the First Billion Years to the Present

The ngVLA will provide a 10x improvement in depth and area for cold gas surveys in galaxies to early cosmic epochs, and will enable routine sub-kiloparsec scale resolution imaging of the gas reservoirs. The ngVLA will afford a unique view into how galaxies accrete and expel gas and how this gas is transformed inside galaxies by imaging their extended atomic reservoirs and circum-galactic regions, and by surveying the physical and chemical properties of molecular gas over the local galaxy population. These studies will reveal the detailed physical conditions for galaxy assembly and evolution throughout the history of the Universe.

### 4.4  KSG4: Using Pulsars in the Galactic Center to Make a Fundamental Test of Gravity

Pulsars in the Galactic Center represent clocks moving in the space-time potential of a super-massive black hole and would enable qualitatively new tests of theories of gravity. They offer the opportunity to constrain the history of star formation, stellar dynamics, stellar evolution, and the magneto-ionic medium in the Galactic Center. The ngVLA's combination of sensitivity and frequency range will probe much deeper into the likely Galactic Center pulsar population to address fundamental questions in relativity and stellar evolution. (Figure 5)

### 4.5  KSG5: Understanding the Formation and Evolution of Stellar and Supermassive Black Holes in the Era of Multi-Messenger Astronomy

The ngVLA will be the ultimate black hole hunting machine, surveying from the remnants of massive stars to the supermassive black holes (SMBHs) that lurk in the centers of galaxies. High-resolution imaging will separate low-luminosity black holes in our local Universe from background sources, providing critical constraints on their formation and growth for all sizes and mergers of black hole-black hole binaries. The ngVLA will also identify the radio counterparts to transient sources discovered by gravitational wave, neutrino, and optical observatories. Its high-resolution, fast-mapping capabilities will make it the preferred instrument to pinpoint transients associated with violent phenomena such as supermassive black hole mergers and blast waves.

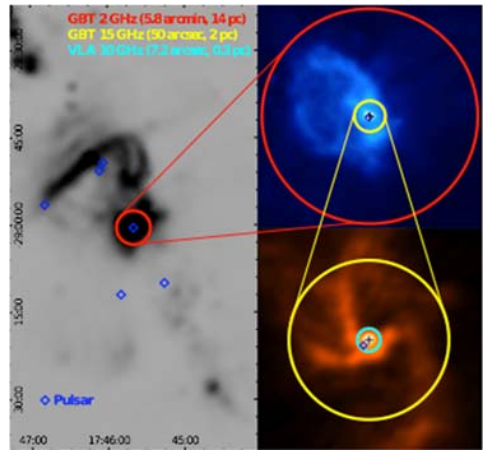

Figure 5. (Credit: R. Wharton) The distribution of known pulsars near the Galactic center. Despite being the region of highest density in the Galaxy and despite having been searched multiple times at a range of frequencies with sensitivities comparable to that of the VLA, only a small number of pulsars are known while up to ~1000 are predicted.

## 5. SCIENCE & TECHNICAL REQUIREMENTS

The Key Science Goals and all other science use cases were parameterized and analyzed [12] to determine the science requirements for the ngVLA [4]. While this aspect of the requirements definition is top-down and mission-driven, some judicious adjustment of the requirements is still appropriate. A primary science requirement for the ngVLA is to be flexible enough to support the breadth of scientific investigations that will be proposed by its creative scientist-users over the decades-long lifetime of the instrument. The requirements have therefore been adjusted to provide a balanced, flexible, and coherent complement of capabilities. The requirements that drive the design are described below.

**Frequency Coverage:** The ngVLA should be able to observe in all atmospheric windows between 1.2 and 116 GHz. These frequency limits are bracketed by spectral line emission from HI and CO respectively.

**Continuum Sensitivity:** A continuum sensitivity of better than 0.02 μJy/bm at 30 GHz and 0.2 μJy/bm 100 GHz is required for studying protoplanetary disks (KSG1). This requires a combination of large collecting area and wide system bandwidth.

**Line Sensitivity:** A line sensitivity of 30 μJy/bm/km/s for frequencies between 10 and 50 GHz is simultaneously required to support both astrochemistry studies and deep/blind spectral line surveys. A line sensitivity of 1 – 100 mK at 0.1" – 5" angular resolution and 1 – 5 km/s spectral resolution between 70 and 116 GHz is required to simultaneously support detailed studies of CO and variations in gas density across the local universe. The spectral line cases push the system design towards quantum-limited noise performance at the expense of bandwidth above 10 GHz.

**Angular Resolution:** A synthesized beam having a FWHM better than 5 mas with uniform weights is required at both 30 and 100 GHz, while meeting the continuum sensitivity targets.

**Largest Recoverable Scale:** Angular scales of >20" x (100 GHz/n) must be recovered at frequencies n < 100 GHz. A more stringent desire is accurate flux recovery on arcminute scales at all frequencies. These scales approach the size of the primary beam of an 18m dish, so both shorter baselines and a total power capability are necessary to completely fill in the central hole in the *(u, v)*-plane.

**Surface Brightness Sensitivity:** The array shall provide high-surface brightness sensitivity over the full range of angular scales recoverable with the instrument. This leads to a centrally condensed distribution of antennas.

**Brightness Dynamic Range:** The system brightness dynamic range shall be better than 50 dB for deep field studies. This requirement pushes a number of systematic requirements including pointing, gain, and phase stability.

**Survey Speed:** The array shall be able to map a ~10 square degree region to a depth of ~1 μJy/bm at 2.5 GHz and a depth of ~10 μJy/bm at 28 GHz within a 10 hr epoch for localization of transient phenomena identified with other instruments. Holding collecting area and receiver noise constant, this favors smaller apertures.

**Beamforming for Pulsar Search, Pulsar Timing and VLBI:** The array shall support no less than 10 beams spread over 1 to 10 subarrays that are transmitted, over the full available bandwidth, to a VLBI recorder/correlator, pulsar search engine or pulsar timing engine. The pulsar search and timing engine must be integral to the baseline design.

**Science Ready Data Products:** The primary data product delivered to users shall be calibrated images and cubes. Uncalibrated/"raw" visibilities shall be archived to permit reprocessing. Producing these higher-level data products requires some standardization of the initial modes/configurations that the system is used in (e.g., limited tuning options), and repeatability/predictability from the analog system to reduce the calibration overheads.

**Cost (Construction & Operations):** The construction cost of the array shall not exceed $1.5B USD (2016) and the annual operation cost of the array shall not exceed $75M USD (2016). The operations cap favors designs with less elements (i.e., small N, large D), which is in conflict with a number of performance metrics.

## 6. SYSTEM CONCEPT

### 6.1 Site Selection and Performance

Decades of data on the quality of the VLA site as an observing location are available, including extensive studies of opacity and phase stability, establishing the Plains of San Agustin as a good site for millimeter-wavelength interferometry [19]. The VLA site was used for acceptance testing of the original ALMA antennas, including observations up to 230 GHz, and the experience was that the VLA site, at 2124 m elevation is a quality 90 GHz site - comparable to the Plateau de Bure site in overall quality [22].

Analysis of data from the VLA site atmospheric phase monitor shows that fast switching phase calibration at 3 mm should be viable, day or night, for most of the year with a 30 sec total calibration cycle. There should be a 25 mJy calibrator source within 2° in 98% of observed fields, ensuring short slews. Such a calibrator is adequate to ensure that the residual rms phase noise due to SNR on the phase calibrator is much less than that due to the troposphere, even for a 30sec cycle time with only 3 sec on the calibrator each visit [19, 20]. The project is also investigating radiometric phase correction techniques as part of the ngVLA project to increase the total phase calibration cycle time.

In addition, the VLA is remote enough that Radio Frequency Interference (RFI) is not a particular problem, in comparison to other sites, so it will also be possible to observe at lower frequencies [21]. The degree of characterization of the site reduces the risk in site selection, and leveraging existing infrastructure could create significant cost savings for both the construction and operation of the array.

Because of the quality of the site for both low- and high-frequency observing, and the existing infrastructure, we choose to center the ngVLA near the current VLA.

### 6.2 Array Configuration

The main array configuration will consist of 214 18m antennas at the approximate locations shown in Figure 1. The array collecting area is distributed to provide high surface brightness sensitivity on a range of angular scales spanning from approximately 1000 to 10 mas (see Table 3). In practice, this means a core with a large fraction of the collecting area in a randomized distribution to provide high snapshot imaging fidelity, and arms extending asymmetrically out to ~1000 km baselines, filling out the *(u, v)*-plane with Earth rotation and frequency synthesis.

The array configuration is practical, accounting for logistical limitations such as topography and utility availability. Investigations are underway to improve the imaging sensitivity and fidelity while accounting for additional limitations such as local RFI sources and land management/availability.

Table 3. Radial Distribution of collecting area.

| Radius | Collecting Area Fraction |
|---|---|
| 0 km < R < 1.3 km | 44% |
| 1.3 km < R < 36 km | 35% |
| 36 km < R < 1000 km | 21% |

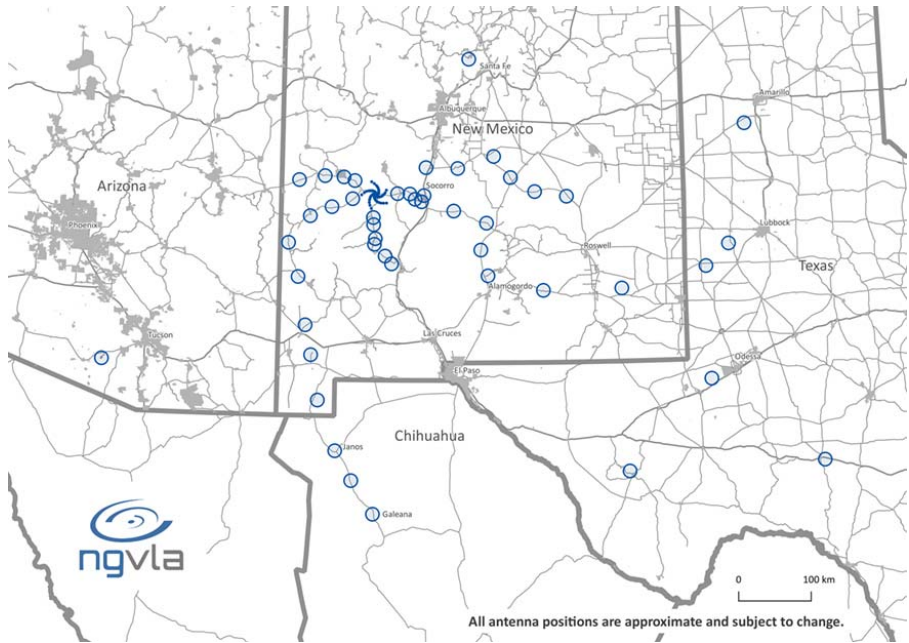

Figure 6. ngVLA Array Configuration Rev. B (Spiral-214). Antenna positions are still notional, but are representative for performance quantification.

The configuration will be a primary area for investigation in the coming years. We have investigated different Briggs weighting schemes for specific science applications [3], and find that the current configuration provides a reasonable compromise and baseline for further iteration.

The design has been extended from the main interferometric array to include both a short spacing array and total power dishes [16]. This is necessary after a review of the key science cases, as these are dependent on the recovery of large scale structure that approaches the size of the antenna primary beam. A cumulative histogram of the minimum baseline required to recover the largest angular scale of interest is shown in Figure 7.

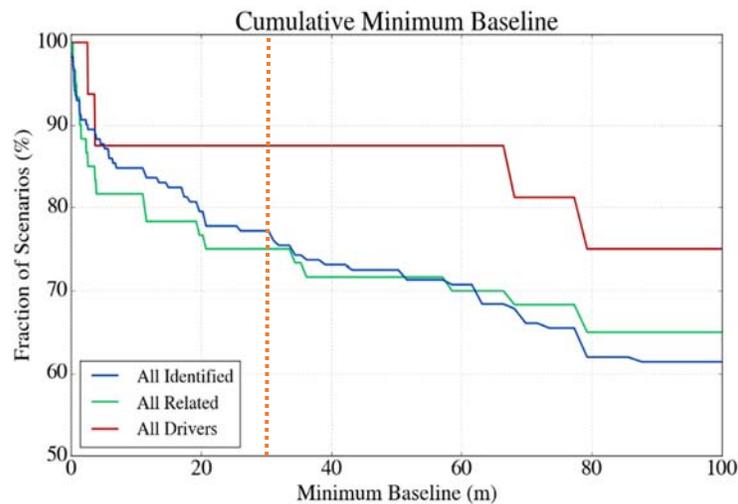

Figure 7. Cumulative histogram of the minimum baseline required to recover the largest angular scale of interest. Approximately 75% of cases can be supported by a 1.65*D spacing between 18m diameter dishes, shown as a dashed vertical line. The remaining 25% of cases require a large single dish, or short baseline array and total power antennas. [12]

An auxiliary short baseline array (SBA) of 19 reflector antennas of 6m aperture will be sensitive to a portion of the larger angular scales undetected by the main array. The SBA will provide spacings from 11m to 60m, providing comparable surface brightness sensitivity to the main array, in equal observing time, when the main array is *(u, v)*-tapered to the natural resolution of the SBA. This allows for commensal observing, and more importantly, full cross-correlation and cross-calibration of the SBA and main array. The array distribution is semi-randomized to improve the point spread function. [15]

The SBA will be combined with 4 18m (main-array) antennas used in total power (TP) mode to completely fill in the central hole in the *(u, v)*-plane left by the 6m dishes. It is a design goal to share the mount design of the 18m interferometric array antennas and the TP antennas, but this will require further study. The array configuration elements are summarized in Table 4.

Table 4. Elements within the ngVLA configuration.

| Array Element | Aperture Diameter | Quantity | $B_{MIN}$ | $B_{MAX}$ | $F_{MIN}$ | $F_{MAX}$ |
|---|---|---|---|---|---|---|
| Main Interferometric Array | 18m | 214 | 30 m | 1005 km | 1.2 GHz | 116 GHz |
| Short Baseline Array | 6m | 19 | 11 m | 56 m | 1.2 GHz | 116 GHz |
| Total Power / Single Dish | 18m | 4 | - | - | 1.2 GHz | 116 GHz |

### 6.3 Array Calibration

The calibration strategy for ngVLA is being developed early in the design so that it may guide the design of the hardware elements. The size and complexity of the calibration and imaging pipeline requires that the system design be responsive to its needs, and it should drive the design where possible.

A secondary concern is the efficiency of the calibration process. Algorithms used must be suitable for parallel processing, antennas must not require much individual attention, and minimal human intervention should be generally required. The calibration overheads applied will vary with the science requirements of a given observation, and less rigorous (and computationally or time efficient) calibration approaches will be applied when possible.

The general calibration strategies under consideration for the reference design are summarized below.

**Fast Atmospheric Phase Calibration**: Rapid atmospheric phase fluctuations will be mitigated by a combination of relative water vapor radiometry (WVR) and antenna switching cycles to astronomical phase calibrators. The switching cycle time will depend on empirical validation of the strategy, but is expected to be necessary on one to ten minute scales. The antenna will be designed to both house the WVR and move 4° on sky and settle to within the pointing specification with 10 seconds for elevation angles <70°. [17]

**Slow Atmospheric & Electronic Phase Calibration**: Slow atmospheric and electronic phase calibration will be achieved by traditional approaches, with astronomical phase calibrator observations bracketing all observations. Several astronomical calibrators may be used to map the slow varying terms, including ionospheric fluctuations.

**Amplitude Calibration**: A list of known astronomical amplitude calibrators will be used to correct for system gain fluctuations within an observation and between observations taken over an extended period of time. The calibration pipeline will maintain a history of recent solutions to enable look-up of prior values.

**Bandpass Calibration**: At a minimum, the system would first correct for digital effects, given the predictable bandpass ripple from FIR filters. The number of setups in the analog portions of the system will be limited, so typical calibration can correct for analog bandpass effects based on historical lookup tables that are updated as the configuration of the system changes (when an antenna is serviced).

**Polarization Calibration**: The use of linear feeds will require polarization calibration for most observations. Feeds will be placed at different (but known) position angles in the various antennas, so a single observation of a point source can solve simultaneously for the polarization leakage terms and the source polarization. Calibration for polarization as a function of position within the antenna beam will be assumed to be time invariant and corrected based on look-up tables.

**Relative Flux Calibration**: This calibration is used to tie together observations of a source taken over an extended period. The system will model opacity based on barometric pressure and temperature monitored at the array core and each outlying station. A temperature stabilized noise diode will provide a reference, and when combined with corrections for modeled atmospheric opacity, we can assume a constant ratio in power from the switched noise calibrator and the source.

**Absolute Flux Calibration**: Absolute flux scale calibration will employ similar methods to relative calibration, with two notable changes. First, atmospheric tipping scans will be used to empirically determine atmospheric opacity. Second, observations of astronomical flux calibrators will be used, along with the switched power system, to determine the absolute flux of the source.

The ngVLA will need to maintain multiple lists of calibrators by calibration intent. The flux calibrator list can be relatively small and based on the one built and maintained by the VLA. An extensive grid of sources will be required for phase and amplitude calibration. The large range of baselines present on the ngVLA means that it cannot be assumed that the source is unresolved, and these calibrators themselves must be imaged before use in the calibration process.

### 6.4 Antenna

The antenna concept strikes a balance between competing science and the programmatic targets for life cycle cost. Sensitivity goals will be met by the total effective collecting area of the array. The reference design includes 214 antennas of 18m aperture (main array) and 19 antennas of 6m aperture (short baseline array) using an offset Gregorian optical design.

The inclusion of frequencies down to 1.2 GHz when combined with the operational cost targets significantly constrain the optical configuration. The use of feeds with wide illumination angles decreases their size such that they can be mounted within shared cryostats. This choice constrains the secondary angle of illumination to a degree that only Gregorian optical designs are practical. However, with a science priority of high imaging dynamic range in the 10-50 GHz frequency range, an offset Gregorian is near optimal. The unblocked aperture will minimize scattering, spillover and sidelobe pickup. Both performance and maintenance requirements favor antenna optical configurations where the feed support arm is on the "low side" of the reflector.

The optimization for operations and construction cost suggests that a smaller number of larger apertures is preferable to larger numbers of small apertures. Survey speed requirements push the opposite direction, and a compromise value of 18m diameter is adopted for the reference design. The design aims for Ruze performance to 116 GHz, with a surface accuracy of 160 μm RMS ($\lambda/16$ @ 116 GHz) for the primary and subreflector combined under precision environmental conditions. The antenna optics are optimized for performance above 5 GHz with some degradation in performance accepted at the lowest frequencies due to diffraction, in exchange for more stiffness in the feed arm to improve pointing performance.

Since the ngVLA is envisioned as a general purpose, PI-driven, pointed instrument (rather than a dedicated survey telescope), the optics will be shaped to optimize the illumination pattern of single pixel feeds, increasing antenna gain while minimizing spillover.

High pointing accuracy will also be necessary to provide the required system imaging dynamic range. With an unblocked aperture, variations in the antenna gain pattern are expected to be dominated by pointing errors. Preliminary requirements are for absolute pointing accuracy of 18 arc-seconds RMS, with referenced pointing of 3 arc-seconds RMS, during the most favorable environmental conditions. [16]

The mechanical and servo design is a typical altitude-azimuth design, Figure 8. Initial studies suggest pedestal designs are expected to have lower life-cycle cost while meeting pointing specifications. The antenna mechanical and servo design will need to be optimized for rapid acceleration and a fast settling time, in order to manage the switching overhead associated with short slews.

The project is presently pursuing a reference design to specifications for the 18m antenna with General Dynamics Mission Systems (GDMS). A parallel study into a composite design concept with the National Research Council of Canada (NRCC) is also underway, and NRCC are also preparing a reference design for the 6m short baseline array antenna. All three costed designs will be delivered in the fall of 2018.

The short baseline array 6m aperture design shares the majority of its specifications with the main antenna, including the interfaces with the front end equipment such that feeds, receivers and other antenna electronics are interchangeable between the two arrays. The design employs a composite reflector and backup structure on a steel pedestal mount. The mount includes space to house the digital electronics, power supplies and servo system, Figure 8.

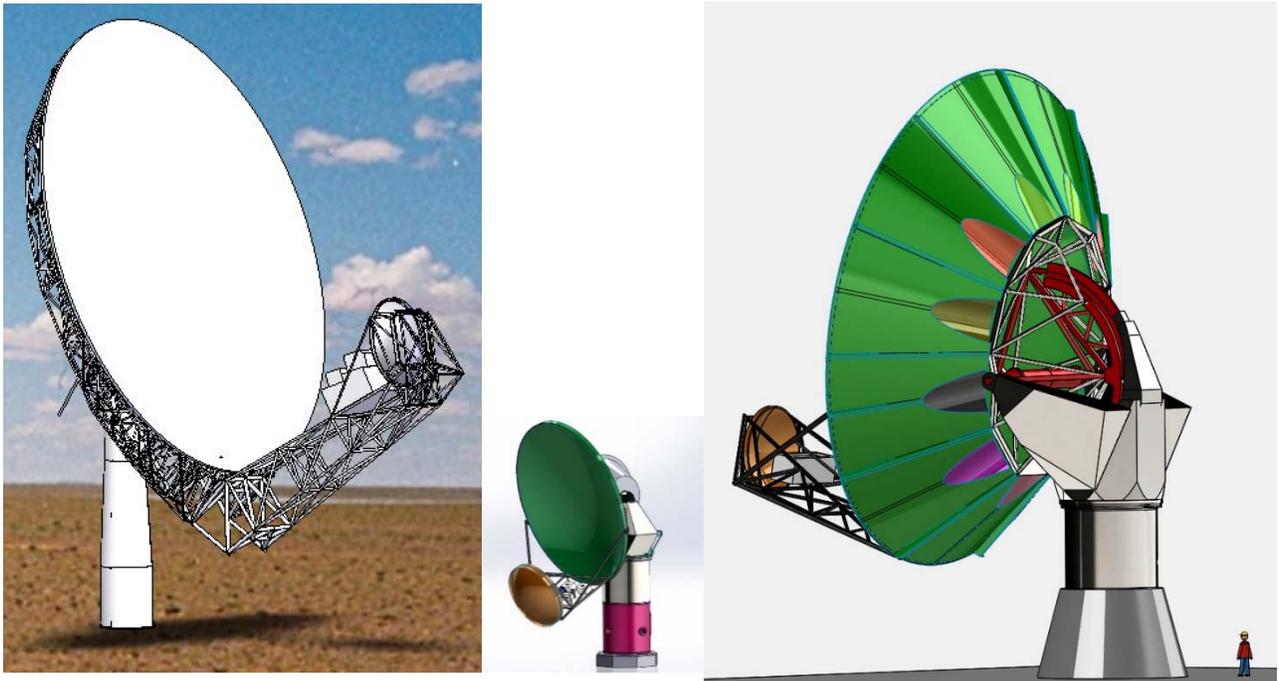

Figure 8. Left: ngVLA 18m antenna reference design concept prepared by GDMS. Center: 6m short spacing array antenna concept prepared by NRCC. Right: ngVLA 18m antenna composite design concept prepared by NRCC.

## 6.5 Receiving Electronics

The ngVLA will provide continuous frequency coverage from 1.2 – 50.5 GHz and 70 – 116 GHz in multiple bands. Receivers will be cryogenically-cooled, with the receiver cryostats designed to integrate multiple receiver bands to the extent possible. Limiting the number of cryostats will reduce both maintenance and electrical power costs. The total number of bands required strongly depends on their fractional bandwidths: maximizing bandwidths will reduce the number of cryostats, with a possible penalty in sensitivity. Feeds for all receiver bands are cooled, and fully contained within the cryostat(s).

The baseline ngVLA receiver configuration consists of the low-frequency receiver (1.2 – 3.5 GHz) in one cryostat, and receivers spanning from 3.5 to 116 GHz in a second cryostat.

Bands 1 and 2 employ wideband feed horns and LNAs, each covering L+S bands, and C+X bands. Quad-ridged feed horns (QRFHs) are used, having dual coaxial outputs. Due to improved optical performance (reducing $T_{SPILL}$), cooled feeds, and the simplified RF design sensing linear polarization, the $T_{SYS}$ is lower than current VLA L, S bands and comparable for C and X bands. Overall aperture efficiency and $T_{SYS}$ are slightly degraded from optimal due to the wider bandwidths spanned, but permits a compact package that can be affordably constructed and operated.

The four high-frequency bands (12.3 – 116 GHz) employ waveguide-bandwidth (~1.67:1) feeds & LNAs, for optimum noise performance. Axially corrugated feed horns with circular waveguide output ensure even illumination over frequency and minimal loss.

Table 5. Key parameters of the baseline receiver concept.

| Band # | $f_L$ GHz | $f_M$ GHz | $f_H$ GHz | BW GHZ | Aperture Eff., $\eta_A$ | | | Spillover, K | | | $T_{RX}$, K | | | $T_{SYS}$, K | | |
|---|---|---|---|---|---|---|---|---|---|---|---|---|---|---|---|---|
| | | | | | @$f_L$ | @$f_M$ | @$f_H$ | @$f_L$ | @$f_M$ | @$f_H$ | @$f_L$ | @$f_M$ | @$f_H$ | @$f_L$ | @$f_M$ | @$f_H$ |
| 1 | 1.2 | 2.0 | 3.5 | 2 | 0.78 | 0.79 | 0.74 | 10 | 7 | 5 | 9 | 10 | 10.5 | **25** | **23** | **22** |
| 2 | 3.5 | 6.6 | 12.3 | 8.8 | 0.78 | 0.79 | 0.70 | 10 | 6 | 4 | 11 | 12 | 15 | **28** | **25** | **26** |
| 3 | 12.3 | 15.9 | 20.5 | 8.2 | 0.84 | 0.87 | 0.86 | 4 | 4 | 4 | 9 | 10 | 11 | **20** | **22** | **31** |
| 4 | 20.5 | 26.4 | 34 | 13.5 | 0.83 | 0.86 | 0.83 | 4 | 4 | 4 | 11 | 13 | 18 | **32** | **33** | **38** |
| 5 | 30.5 | 39.2 | 50.5 | 20 | 0.81 | 0.82 | 0.78 | 4 | 4 | 4 | 17 | 21 | 27 | **36** | **45** | **105** |
| 6 | 70 | 90.1 | 116 | 46 | 0.68 | 0.61 | 0.48 | 4 | 4 | 4 | 39 | 41 | 60 | **116** | **65** | **181** |

(*)Assumes 1mm PWV for band 6, 6mm PWV for others; 45 deg elev. on sky for all.

The electronics concept relies on integrated receiver packages [13] to further amplify the signals provided by the cryogenic stage, down convert them if necessary, digitize them, and deliver the resultant data streams by optical fiber to a moderately remote collection point (typically the antenna pedestal) where they can be launched onto a conventional network for transmission back to the array central processing facility. Interfaces are provided for synchronization of local oscillators (LO's) and sampler clocks, power leveling, command and control, health and performance monitoring, and diagnostics for troubleshooting in the event of component failure.

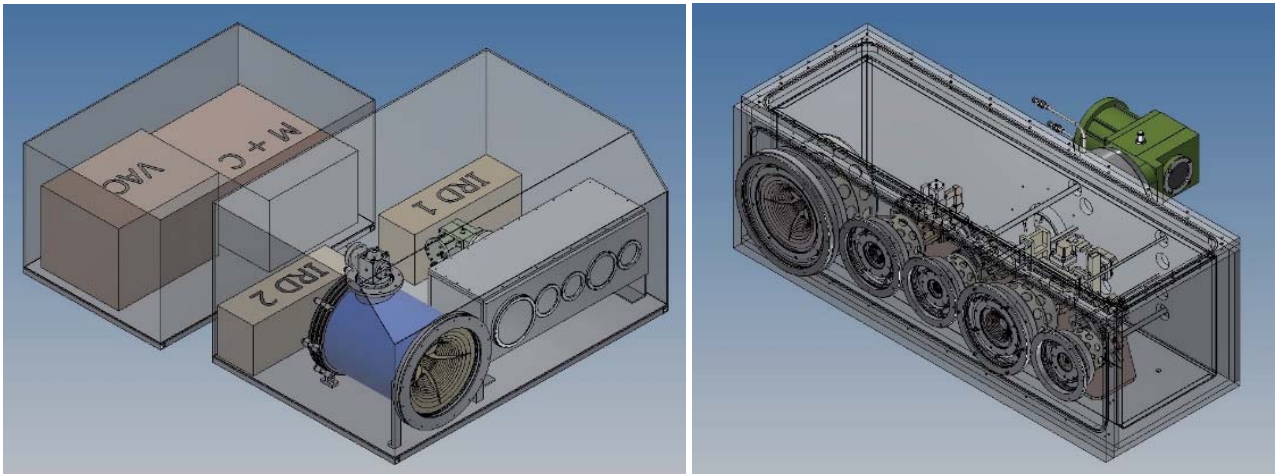

Figure 9. Front end component packaging at the secondary focus of the antenna. Band selection and focus are achieved with a dual-axis translation stage. The integrated receiver packages (labeled IRD 1 and IRD 2) are located in close proximity to the cryostats. Bands 2-6 are housed within in single cryostat.

The integrated receiver concept is central to the antenna electronics concept for the ngVLA. Compact, fully-integrated, field-replaceable, warm electronic modules support single-stage, direct-to-baseband downconversion (when needed), followed by a very low-power, low-overhead digitization scheme and an industry-standard fiber optic interface carrying unformatted serial data. The frequency plan is shown in Figure 10. [13]

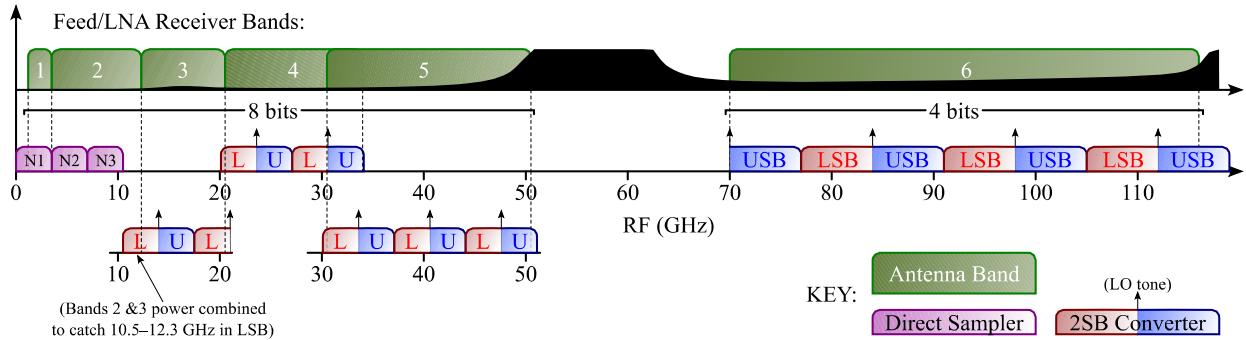

Figure 10. Present sampling concept employing integrated receiver technology for both direct and dual sideband converter/samplers.

## 6.6 Reference Distribution & Data Transmission

Given the large extent of the array, multiple time and frequency reference distribution concepts will likely be required to optimize for cost and performance. The array will likely be built as a combination of two different design methodologies.

A large number of antennas located on the Plains of San Agustin, would be part of a compact core, and each antenna in the core would be connected directly to a central processing facility by fiber optics. Over 80% of the ngVLA antennas will be within this region. Clocks and local oscillator signals may be generated locally at the antenna and locked to a central reference with round trip phase correction, possibly sharing a DTS transceiver (Figure 11).

The remainder of the antennas, the long baseline antennas, would fall into a VLBI station model with a number of local oscillator (LO) and data transmission stations located beyond the central core. These stations will be linked to the central timing system, correlator, and monitor and control system via long haul fiber optics. Several options will be explored for precision timing and references at these stations, including local GPS-disciplined masers, fiber optic connections to the central site, and satellite-based timing.

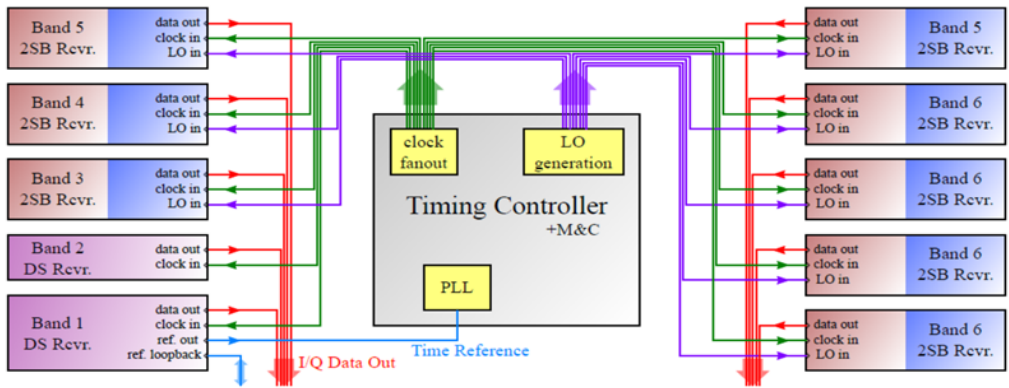

Figure 11. Schematic of clock and LO distribution inside the antenna.

## 6.7 Central Signal Processor

The CSP ingests the voltage streams recorded and packetized by the antennas and transmitted via the data transmission system, and produces a number of low-level data products to be ingested by the archive. In addition to synthesis imaging, the CSP will support other capabilities required of modern telescopes to enable VLBI and time-domain science. The functional capabilities of the CSP include:

- Auto-correlation

- Cross-correlation
- Beamforming
- Pulsar Timing
- Pulsar Search
- VLBI Recording

The CSP data products will vary by operation mode. The most common will be raw/uncalibrated visibilities, recorded in a common data model. The CSP will include all necessary "back end" infrastructure to average visibilities and package them for the archive, where they will be recorded to disk in a standard format. Calibration of these data products will be the responsibility of asynchronous data post-processing pipelines that are outside the scope of the CSP element.

The CSP will support multiple sub-arrays. A key requirement for the system is the degree of commensality supported both within a sub-array and by commensal sub-arrays. The following commensal modes will be supported:

- Cross-correlation and Auto-correlation (Commensal within a sub-array)
- Cross-correlation & Pulsar Timing (Commensal sub-arrays)
- Cross-correlation & Pulsar Search (Commensal sub-arrays)
- Cross-correlation & VLBI Recording (Commensal sub-arrays)

Providing correlation and beamforming products simultaneously within a sub-array is also under evaluation. Such a mode would aid calibration of the beamformer, and provide for localization/imaging concurrent with time-domain observations. The degree of commensality is expected to be a cost/complexity driver in the system and will be optimized on a best value basis.

The ngVLA correlator will employ an FX architecture, and will process an instantaneous bandwidth of up to 20 GHz per polarization. The correlator-beamformer Frequency Slice Architecture [14] developed by NRC Canada for the SKA Phase 1 CSP Mid Telescope is well suited to ngVLA demands and is under evaluation for the reference design. This architecture will scale to the additional ngVLA apertures, bandwidth, and commensal mode requirements. Adopting this architecture will significantly reduce the non-recurring engineering costs during the design phase, while additional improvements in electrical efficiency can be expected from one additional FPGA manufacturing process improvement cycle due to ngVLA's later construction start date. Key performance requirements for the correlator are summarized in Table 6.

Table 6. Correlator-beamformer key specifications.

| Requirement Description | Specification |
|---|---|
| Number of Connected Antennas | 256 total, composed of 214 18m elements, 19 6m elements, 12 VLBI stations (TBC), and 11 spare inputs |
| Maximum Baseline Length | 1,000 km |
| Maximum Instantaneous Bandwidth | 20 GHz |
| Maximum Number of Channels | 300,000 channels |
| Highest Frequency Resolution | 400 Hz, corresponding to 0.1 km/s resolution at 1.2 GHz. |
| Pulsar Search Beamforming | ≥91 beams, 60 km diameter sub-array, 1" coverage |
| Pulsar Timing Beamforming | ≥10 independent sub-arrays, 50 beams total |

## 6.8 Post Processing System

The software architecture for ngVLA will leverage NRAO's existing algorithm development in reducing VLA and ALMA data and the CASA software infrastructure. The array will have a progressive series of data products suitable to different users groups. The data products may also change based on how well supported a mode is – common modes should have higher level data products that add value to the user, while clearly not all permutations can benefit from such a degree of automation.

As with the VLA, the fundamental data product that will be archived are uncalibrated visibilities. The online software system will also produce flags to be applied to the visibilities that would identify known system problems such as antennas being late on source, or the presence of RFI.

Automated post-processing pipelines will calibrate the raw data and create higher-level data products (typically image cubes) that will be delivered to users via the central archive. Calibration tables that compensate for large-scale instrumental and atmospheric effects in phase, gain, and bandpass shapes will be provided. Data analysis tools will allow users to analyze the data directly from the archive, reducing the need for data transmission and reprocessing at the user's institution.

The VLA and ALMA "Science Ready Data Products" project will be an ngVLA pathfinder to identify common high-level data products that will be delivered to the Principal Investigator and to the data archive to facilitate data reuse. This will also enable the facility to support a broader user base, possibly catering to astronomers who are not intimately aware of the nuances of radio interferometry, thereby facilitating multi-wavelength science.

## 7. CONCLUSION

The ngVLA will open a new window on the Universe through ultra-sensitive imaging of thermal line and continuum emission down to milliarcsecond resolution, as well as unprecedented broadband continuum polarimetric imaging of non-thermal processes.

The array collecting area is distributed to provide high surface brightness sensitivity on a range of angular scales spanning from approximately 1000 to 10 mas. In practice, this means there is a core with a large faction of the collecting area in a randomized distribution to provide high snapshot imaging fidelity, and arms extending asymmetrically out to ~1000 km baselines, filling out the *(u, v)*-plane with Earth rotation and frequency synthesis.

The main interferometric array is a homogeneous array of 214 18m apertures. The selection of the number of apertures and the antenna diameter is supported by parametric cost and performance estimation. Given the array extent, the antenna locations are fixed, with no reconfiguration capability. All antennas are outfitted with six front ends that provide access to the atmospheric windows spanning 1.2 GHz to 50.5 GHz and 70 GHz to 116 GHz.

In addition to the main interferometric array, a short baseline array of 19 6m apertures is included along with four 18m total power antennas. The total power antennas operate as single dishes or elements of the main interferometric array depending on the operational requirements. The size of the short spacing array and total power array complement the surface brightness sensitivity of the 18m interferometric array on overlapping angular scales, ensuring that large scale extended emission and structure is accurately recovered.

The ngVLA requirements, concept and design are maturing. The reference design will soon be sufficiently defined to estimate the performance, cost, and technical risk associated with the design.

**CITATION FORMAT**

Selina, et. al., "The Next-Generation Very Large Array: A Technical Overview", SPIE Astronomical Telescopes & Instrumentation, AS18, 10700-55, (2018); DOI.

**COPYRIGHT NOTICE:**